\documentstyle[preprint,aps,prb]{revtex}

\begin{document}
%\preprint{HEP/123-qed}
\title{Superdiffusion in Decoupled Continuous Time Random Walks}
\author{C. Budde$^1$, D. Prato$^1$ and M. R\'{e}$^1$}
\address{$^1$Facultad de Matem\'{a}tica, Astronom\'{\i}a y F\'{\i}sica,
Universidad Nacional de\\
C\'{o}rdoba, Ciudad Universitaria, 5010 C\'{o}rdoba, Argentina\\
prato@mail.famaf.unc.edu.ar \\}
\date{today}
\maketitle

\textwidth 15truecm
\oddsidemargin 0.5truecm

\begin{abstract}

Continuous time random walk models with decoupled waiting time density are
studied.
When the spatial one jump probability density belongs to the Levy
distribution type and the total
time transition is exponential a generalized superdiffusive regime is established.
This is verified by showing that the square width of the probability
distribution (appropriately defined)grows 
as $t^{2/\gamma}$ with $0<\gamma\leq2$ when  $t\rightarrow \infty$.
An important connection of our results and those of Tsallis' nonextensive
statistics  
is shown. The normalized q-expectation value of $x^2$ calculated with the
corresponding probability distribution behaves exactly as $t^{2/\gamma}$ in
the asymptotic limit.

\end{abstract}

\newpage

Recently there has been an increasing interest in dynamical processes that
display
anomalous diffusion \cite {Tsallis,Prato,Zasla,Klafter,Blumen,Zumofen}. These
phenomena have been 
characterized by a non linear time dependence of the mean square displacement of the walker 
\begin{equation}
\langle x^{2}(t)\rangle \;\sim\;t^{\alpha}
\end{equation}
with $\alpha\;\neq 1$ ($\alpha = 1$ gives normal diffusion) since $\langle x^2\rangle $ is an usual estimator of the square width of the probability distribution at time $t$.  In this way, for anomalous diffusion we have that the probability distribution width grows faster (slower) for $\alpha > 1$ ($\alpha <1$) than it does for normal diffusion. Examples of Eq.(1) with $\alpha\; < \;1$ (dispersive
diffusion or subdiffusion) are found in disordered
media like glasses and fractals structures \cite
{Scher,Shles,Alexa,Blumen2}.
Chaotic dynamics and turbulence, on the other hand, give rise to enhanced
diffusion (superdiffusion), i.e., $\alpha >1$
\cite{Monin,Batchelor,Grossmann,Shles2}.
Diffusion like behaviour can be modeled in the framework of the
continuous time 
random walk (CTRW), where the central magnitude is the waiting time density (WTD), $\Psi(x,t)$, as was first proposed by Montroll and Weiss \cite{Montroll}.  In this way $\Psi(x,t)\, dx dt$ represents the probability that the
walker makes a jump of length $x$ after a time $t$ since the last jump.

When it exists the walker mean square displacement from the
  origin, at time 
$t$, can be calculated from the Fourier transform of the spatial probability density $P\left( x,t\right) $ according to
\begin{equation}
<x^{2}(t)>\;=\;\int x^{2}\;P(x,t)\;dx\;=\;-\frac{\partial^{2}
P(k,t)}{\partial k^{2}}|_{k=0}
\end{equation}
where $P(x,t)\, dx$ is the conditional probability of finding the walker between $x$ and $x+dx$
 at time $t$ 
given that it started at $x=0$ at time $t=0$, and $P(k,t)$ represents its
Fourier transform in  the space variable.  As was shown in previous papers, it is imperative to use a coupled WTD in order to have a finite step mean square displacement in the superdiffusive regime {\it i.e.} when the distribution width grows faster than in normal diffusion.

This requirement can be understood since for a decoupled WTD
\begin{equation}
\label{e2}
\Psi(x,t)\;=\;p(x)\;\psi(t)
\end{equation} 
a divergent mean square displacement for the one step distribution will make infinite $\langle x^2\left( t\right) \rangle $ also.  This would suggest a formally infinite diffusion coefficient as pointed out by Ott {\it et al.} \cite{ott}, while the experimental results for elongated micelles analyzed in their paper let them define ``a (distance-dependent) diffusion constant´´. Nevertheless the coupled WTD requirement is not necessary if we consider alternative estimators of the width fo the probability distribution when $\langle x^2\left( t\right) \rangle $ diverges, as we will do in this letter.

In particular it is well known that the Levy family of probability densities \cite{Montroll2}, has divergent second moment. { \it In spite
of this divergence of their second moment a finite width can still be defined
for these distributions by considering alternative estimators, such as the inverse of $P\left( x=0,t\right) $ or the width at half of the maximun value}.  Notice that with any of these criteria we get a linear law for the time dependence of the square width in normal diffusion as we do with $\langle x^2\left( t\right) \rangle $. One of the most familiar examples of Levy distributions is the Lorentz or
Cauchy distribution. This distribution arises naturally in the study of the
line shape of fluctuation spectra of stochastic Markov processes \cite{van
Kampen}. It is known that the Fourier transform of Levy distributions are,
by definition, given by
\cite{Montroll2}

\begin{equation}
p(k)\;=\;\exp(-c|k|^{\gamma})\;\;, \;\;\;0<\gamma\leq2 \label{Levy}
\end{equation}

In the framework of a non-extensive thermostatistics, Tsallis and
collaborators
 \cite{Tsallis,Prato} have obtained, from a variational principle,
 probability 
densities that share with the Levy family of density functions the asymptotic behavior for
large steps.
The densities obtained in these papers are  (with appropriate simple constraints) the extremal
of a functional entropy \cite{tsallis88} $S_{q}$   where the parameter $q\;<\;3$ identifies a particular density.  Tsallis' formulation generalizes the classical Boltzmann-Gibbs 
(obtained for the particular value $q\;=\;1$) variational formulation.  The relationship between parameter $q$
and 
the exponent $\gamma$ are related by $\gamma\;=\; \frac{3-q}{q-1}\;$ in the range
$\;\frac{5}{3}
\;\leq\; q \;< 3\;$, where the second moment diverges, while $\;\gamma\;=\;2\;$ for $\;q<\frac{5}{3}\;$. They
also show \cite{Tsallis,Prato} that 
the  N-jump process constructed from these densities scales as $N^{2/\gamma}$ for $N \rightarrow
\infty$.

Our purpose here is to extend the Tsallis' results to a CTRW with a 
decoupled  WTD. One important result of this extension is that in 
case that the one step probability $p(x)$ (\ref{e2}) is of the Levy type (\ref{Levy}), the width of the conditional probability density $P(x,t)$ grows as $t^{1/\gamma}$ for
$t\rightarrow \infty$, so that for $\gamma <2$ we obtain a superdiffusive regime.

Let us start by re-obtaining Tsallis' results for the N-jump distribution
for a Levy 
flight. The probability density for the walker position after $N$ steps is given by the N-fold convolution of the one-jump distribution
\begin{equation}
P(x,N)\;=\;p(x)*p(x)*.....*p(x)
\end{equation}
whose Fourier transform is 
\begin{equation}
P(k,N)\;=\;[p(k)]^{N}\;=\;\exp(-cN|k|^{\gamma})
\end{equation}
by making use of  Eq. \ref{Levy}.
Anti-transforming the last equation we finally obtain
\begin{eqnarray}
P(x,N)\;&=&\; \frac{1}{2 \pi} \int_{-\infty}^{\infty} dk
\exp(-cN|k|^{\gamma})
 \exp(ikx) \nonumber \\ 
\label{Nconv}\\
&=&\; N^{-1/\gamma} \;\;p(N^{-1/\gamma}\;x)   \nonumber
\end{eqnarray}
where a change of variable has been made.\\

Eq. (\ref{Nconv}) is valid for all N and shows that the width of the N-jump 
distribution, when appropriately defined, is that of the one-jump distribution times the factor
$N^{1/\gamma}$. The 
prefactor ensures the normalization condition of the probability density. As it is readily verified, estimators such as the inverse of $P\left( x=0,t\right) $ or the width at half of the maximun value, which are finite for these distributions, exhibit the $N^{1/\gamma }$ growth. In the CTRW scheme a similar
result may 
be obtained when the transition time probability density is
\begin{equation}
\psi(t)\;=\;\lambda \exp(-\lambda\;t) \label{Poisson}
\end{equation}

In this case 

\begin{equation}
P(x,t)\;=\; \frac{1}{2 \pi} \int_{-\infty}^{\infty} dk
 \; \exp[-\lambda\;t\;(1-p(k))]\;\exp(ikx) \label{CTRW}
\end{equation}

satisfies the Master equation:

\begin{equation}
\frac{\partial P(x,t)}{\partial t}\;=\;\lambda\int_{-\infty}^{\infty} dx'
\;p(x-x')\;P(x',t)\;\;-\lambda\ P(x,t)       \label{eqMa}
\end{equation}
and may be expressed as a power series in time as follows:

\begin{equation}
P(x,t)\;=\;\exp(-\lambda\;t)\delta(x)\;\;+\;
\exp(-\lambda\;t)\;\sum_{n=1}^{\infty}\;
[\;\frac{(\lambda\;t)^{n}}{n!}\;\;  p(n^{-1/\gamma}x)\;]  \label{sol}
\end{equation}
where $\delta(x)$ is Dirac's delta.

In the limit $\lambda\;t \gg\;1$ only small values of $k$ contribute to 
the integral in Eq. (\ref{CTRW}) \cite{Montroll2}, giving rise to 

\begin{equation}
P(x,t)\;\sim \; \frac{1}{2 \pi} \int_{-\infty}^{\infty} dk
 \; \exp(-\lambda\;t\;c|k|^{\gamma})\;\exp(ikx) \;\;\;\; \mbox{for} \;\;\;
 \lambda\;t \gg\;1 \label{CTRW1}
\end{equation}

By making a change of variables we get

\begin{equation}
P(x,t) \;\sim\;(\lambda\;t)^{-1/ \gamma}\;p[(\lambda\;t)^{-1/ \gamma}\;x]
\label{CTRW2}
\end{equation}
then the width of the probability distribution at time $t$ grows as
$(\lambda\;t)^{1/
\gamma}$ in a similar way as was obtained for the discrete time case. As we have mentioned before in normal diffusion $<x^{2}(t)>\;
\sim\;t $, i.e., 
the square width grows as  $t$, while, in the N-jump Levy process, as we
have shown, 
the square width behaves as $N^{2/\gamma}$, and as $t^{2/\gamma}$ in the
continuous time 
case. Since $0<\gamma\leq2 $ we conclude that a superdiffusive regime is
established {\it even}
in the separable CTRW case when the one-jump space distribution belongs to
the Levy family.

In Figure 1 we show a plot of the width of the probability distribution 
versus t (in logarithmic scale) for various values of the parameter
$\gamma$. Here two 
criteria have been used to define the width of the distributions; one of
them just consists in taking the inverse
of $P(x=0,t)$, and the other one is the width taken at half of the maximun
value of the 
distribution. We observe that, for $\lambda t \geq 10$, the width grows
linearly 
with $t^{1/\gamma}$, according to our prediction.

It is worthwhile to notice that our results may be connected with those of
Tsallis
and collaborators obtained in the framework of nonextensive statistics
\cite{Tsallis,Prato}. 
Followig these authors, we calculate the normalized q-expectation value of
$x^{2}$ 
for a N-jump process

\begin{equation}
<x^{2}>_{q}(N)\;=\;\frac{\int_{-\infty}^{\infty}
dx\;x^{2}\;[P(x,N)]^{q}}{\int_{-\infty}^{\infty}
dx\;[P(x,N)]^{q}} \label{q-expN} 
\end{equation}
  
with the following constraint imposed on the single step distribution
\begin{equation}
\sigma^{2}\;=\;\frac{\int_{-\infty}^{\infty}
dx\;x^{2}\;p(x)^{q}}{\int_{-\infty}^{\infty}
dx\;p(x)^{q}} \label{q-exp1}
\end{equation}

Using Eq. (\ref{Nconv}) and performing a change of variable we obtain

\begin{equation}
<x^{2}>_{q}(N)\;=\;N^{2/\gamma}\;\sigma^{2} \label{disp}
\end{equation}

This result may be extended to the continuous time case for 
$\lambda\;t\;\gg 1$

\begin{equation}
<x^{2}>_{q}(t)\;=\;(\lambda\;t)^{2/\gamma}\;\sigma^{2} \label{dispcont}
\end{equation}

The last two  results suggest that the normalized q-expectation values of
$x^{2}$ can be used as an estimator of the square width of the Levy type
distributions obtained in \cite{Prato}, whose normal ($q\;=\;1$) expectation value of $x^{2}$ (second moment) diverges.

To conclude, let us emphasize the main result of this letter: a CTRW, with a
separable waiting time density $\Psi(x,t)\;=\;p(x)\;\psi(t)$, whose one-jump 
space distribution $p(x)$ belongs to a Levy family, gives rise to a generalized 
superdiffusive 
process in the case of an exponential form, Eq. (\ref{Poisson}), for the
total transition 
probability density $\psi(t)$. If, however,  $\psi(t)$ has a long time tail
the above 
conclusion may not be true; work along this line is in progress. Finally,
the non-extensive thermo statistics seems to be the 
appropriate framework to tackle anomalous diffusion.\\\\

AKNOWLEDGMENTS

This work was supported by Grants from CONICET-PID 4892 (1998), CONICOR-PID
4624 and
4643 (1998) and SECYT.UNC-PID 163/99 (2000).

\newpage

Figure caption

Figure 1:  Width of the probability distribution $P(x,t)$ defined in two
ways:
$W_{1}$ = width at the half height of $P(x,t)$ and $W_{2}$ =$1/ [P(x=0,t]$
for various 
one-step Levy distributions determined by the values of $\gamma$ (see text).
Notice the 
linear behavior of the width, with both criteria, in the asymptotic limit
$\lambda t\rightarrow \infty$. The slopes of the curves are consistent with
the 
value $1/ \gamma$ predicted by the theory.

\end{document}